\documentclass[a4paper,twoside,12pt]{article}
\input{obsjkt.sty}

\newcommand{\Msun}{\ensuremath{~{\rm M}_\odot}}                   
\newcommand{\Rsun}{\ensuremath{~{\rm R}_\odot}}                   
\newcommand{\rhosun}{\ensuremath{~\rho_\odot}}                    
\newcommand{\Teff}{\ensuremath{T_{\rm eff}}}                      
\newcommand{\logg}{\ensuremath{\log g}}                           
\newcommand{\EBV}{\ensuremath{E(B\!-\!V)}}                        
\newcommand{\degr}{\ensuremath{^\circ}}                           
\renewcommand{\kms}{~km~s$^{-1}$}                                 
\newcommand{\as}{\ensuremath{^{\prime\prime}}}                    
\newcommand{\chir}{\ensuremath{\chi_\nu^{\,2}}}                   
\newcommand{\hip}{\textit{Hipparcos}}                             
\newcommand{\gaia}{\textit{Gaia}}                                 
\newcommand{\targ}{HP~Dra}
\newcommand{\targfull}{HP~Draconis}

\newcommand{\Msunnom}{\hbox{$\mathcal{M}^{\rm N}_\odot$}}
\newcommand{\Rsunnom}{\hbox{$\mathcal{R}^{\rm N}_\odot$}}
\newcommand{\Lsunnom}{\hbox{$\mathcal{L}^{\rm N}_\odot$}}

\usepackage{rotating}

\begin{document} 

\OBSheader{Rediscussion of eclipsing binaries: \targ}{J.\ Southworth}{2025 October}

\OBStitle{Rediscussion of eclipsing binaries. Paper XXVI. \\ The F-type long-period system HP~Draconis}

\OBSauth{John Southworth}

\OBSinstone{Astrophysics Group, Keele University, Staffordshire, ST5 5BG, UK}


\OBSabstract{\targ\ is a well-detached eclipsing binary containing two \mbox{late-F} stars on an orbit with a relatively large period of 10.76~d and a small eccentricity of 0.036. It has been observed in 14 sectors using the Transiting Exoplanet Survey Satellite (TESS). We use these data plus literature spectroscopic measurements to establish the properties of the component stars to high precision, finding masses of $1.135 \pm 0.002$\Msun\ and $1.098 \pm 0.002$\Msun\ and radii of $1.247 \pm 0.005$\Rsun\ and $1.150 \pm 0.005$\Rsun. We find a much smaller third light than previous analyses, resulting in significant changes to the measured radii. These properties match theoretical predictions for an age of 3.5~Gyr and a solar metallicity. We present a spectrum of the Ca H and K lines in which chromospheric activity is visible from both components. The distance we find to the system, $77.9 \pm 1.2$~pc, matches the \gaia\ DR3 parallax value of $79.2 \pm 0.3$~pc.}


\section*{Introduction}

This work continues our series \cite{Me20obs} of reanalyses of known detached eclipsing binary systems (dEBs) using new photometric observations primarily from the NASA Transiting Exoplanet Survey Satellite \cite{Ricker+15jatis} (TESS). The aim is to use space-based data \cite{Me21univ} to improve the measurements of the properties of the component stars and to add them to the Detached Eclipsing Binary Catalogue \cite{Me15aspc} (DEBCat\footnote{\texttt{https://www.astro.keele.ac.uk/jkt/debcat/}}).

In this work we present a study of \targfull\ (Table~\ref{tab:info}), a late-F-type system with an eccentric orbit of relatively long period. Its variability was first noticed in photometry from the \hip\ satellite \cite{ESA97}, with a period of 6.693~d, and it was given its variable-star designation by Kazarovets et al.\ \cite{Kazarovets+99ibvs}. Its correct orbital period of 10.762~d was established by Kurpi\'nska-Winiarska et al.\ \cite{Kurpinska+00ibvs} and refined by Milone et al.\ \cite{Milone+05aa}.

Milone et al.\ \cite{Milone+05aa} presented an analysis of \targ, using radial velocities (RVs) from the Ca infrared triplet and the \hip\ light curve to simulate the then-expected performance of the \gaia\ mission. This work was updated by Milone, Kurpi{\'n}ska-Winiarska \& Oblak \cite{Milone++10aj} (hereafter MKO10) using additional RVs and $BV$ photometry from Cracow \cite{Kazarovets+99ibvs}. This analysis resulted in measurements of the stellar masses and radii to 1\% precision. The previous finding of apsidal motion \cite{Milone+05aa} became only marginally significant in this analysis.

MKO10 found that 10\% of the light in the system was produced by a source other than the two eclipsing stars, and noted that this was more than could be provided by nearby resolved stars contaminating the photoelectric photometry. They also presented cross-correlation functions in which no trace of a putative third component was visible. Possible explanations are that the additional light comes from an object with few spectral lines (e.g.\ a white dwarf) or from two or more stars none of which are individually identifiable in the spectra. 

Jalowiczor et al.\ \cite{Jalowiczor+21rnaas} identified the object Gaia DR2 2144465183642117888 as a white dwarf companion to \targ, with a common proper motion and an angular distance of 14.405\as. As it is fainter by approximately 10~mag, it contributes a negligible amount of light to the TESS light curve and thus is not responsible for the third light found by MKO10.

Baroch et al.\cite{Baroch+21aa} included \targ\ in their sample of dEBs expected to show apsidal motion dominated by the general-relativistic contribution. They analysed the first three sectors of TESS data but found no clear evidence for apsidal motion.

\begin{table}[t]
\caption{\em Basic information on \targfull. 
The $BV$ magnitudes are each the mean of 94 individual measurements \cite{Hog+00aa} distributed approximately randomly in orbital phase. 
The $JHK_s$ magnitudes are from 2MASS \cite{Cutri+03book} and were obtained at an orbital phase of 0.89. \label{tab:info}}
\centering
\begin{tabular}{lll}
{\em Property}                            & {\em Value}                 & {\em Reference}                      \\[3pt]
Right ascension (J2000)                   & 18 54 53.481                & \citenum{Gaia23aa}                   \\
Declination (J2000)                       & +51 18 29.79                & \citenum{Gaia23aa}                   \\
Henry Draper designation                  & HD 175900                   & \citenum{CannonPickering22anhar}     \\
\textit{Hipparcos} designation            & HIP 92835                   & \citenum{ESA97}                      \\
\textit{Tycho} designation                & TYC 3552-394-1              & \citenum{Hog+00aa}                   \\
\textit{Gaia} DR3 designation             & 2144465183642116864         & \citenum{Gaia21aa}                   \\
\textit{Gaia} DR3 parallax (mas)          & $12.6153 \pm 0.0516$        & \citenum{Gaia21aa}                   \\          
TESS\ Input Catalog designation           & TIC 48356677                & \citenum{Stassun+19aj}               \\
$B$ magnitude                             & $8.54 \pm 0.02$             & \citenum{Hog+00aa}                   \\          
$V$ magnitude                             & $7.93 \pm 0.01$             & \citenum{Hog+00aa}                   \\          
$J$ magnitude                             & $6.853 \pm 0.020$           & \citenum{Cutri+03book}               \\
$H$ magnitude                             & $6.616 \pm 0.016$           & \citenum{Cutri+03book}               \\
$K_s$ magnitude                           & $6.565 \pm 0.017$           & \citenum{Cutri+03book}               \\
Spectral type                             & F9~V + F9~V                 & \citenum{Milone+05aa}                \\[3pt]
\end{tabular}
\end{table}


\section*{Photometric observations}


\targ\ has been observed by TESS in 14 sectors (14, 15, 26, 40, 41, 53, 54, 55, 59, 74, 75, 80, 82, 86). In all cases data are available at 120~s cadence, and these were used for our analysis below. Lower-cadence observations (200, 600 and/or 1800~s) are also available for all sectors but were not used due to their poorer time resolution. The data were downloaded from the NASA Mikulski Archive for Space Telescopes (MAST\footnote{\texttt{https://mast.stsci.edu/portal/Mashup/Clients/Mast/Portal.html}}) using the {\sc lightkurve} package \cite{Lightkurve18}. 

We used the simple aperture photometry (SAP) light curves from the SPOC data reduction pipeline \cite{Jenkins+16spie} for our analysis, and removed low-quality data using the {\sc lightkurve} quality flag ``hard''.  The data were converted into differential magnitude and the median magnitude was subtracted from each sector for convenience. 

\begin{figure}[t] \centering \includegraphics[width=\textwidth]{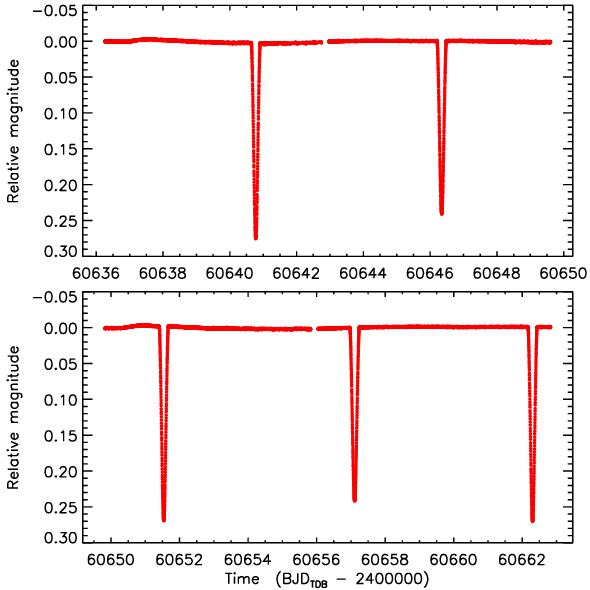} \\
\caption{\label{fig:time} TESS sector 86 photometry of \targ. The flux measurements have 
been converted to magnitude units after which the median was subtracted. The other 13 
sectors used in this work are very similar so are not plotted.} \end{figure}

Fig.~\ref{fig:time} shows the light curve from sector 86; the remaining sectors are similar so are not plotted. Some variability is visible in sector 86 and others (e.g.\ at times around BJD 2460637.5 and 2460651.0). This variability recurs on the orbital period of TESS so is an instrumental signal, not astrophysical.

\section*{Light curve analysis}

The components of \targ\ are well-separated and the light curve is suitable for analysis using the {\sc jktebop}\footnote{\texttt{http://www.astro.keele.ac.uk/jkt/codes/jktebop.html}} code \cite{Me++04mn2,Me13aa}. The profusion of data, the possibility of apsidal motion, and the expected change of third light with TESS sector, meant it was most efficient to model the light curve from each sector individually. The system is in eclipse for only approximately 10\% of the time, so we removed data away from an eclipse to save computational time. This was done by detecting each fully-observed eclipse and retaining all data during eclipse plus an additional 0.1~d both before and after. Each eclipse was then normalised to zero differential magnitude by fitting and subtracting a straight line to the out-of-eclipse data, in order to remove slow variations of either instrumental or astrophysical origin.

\begin{figure}[t] \centering \includegraphics[width=\textwidth]{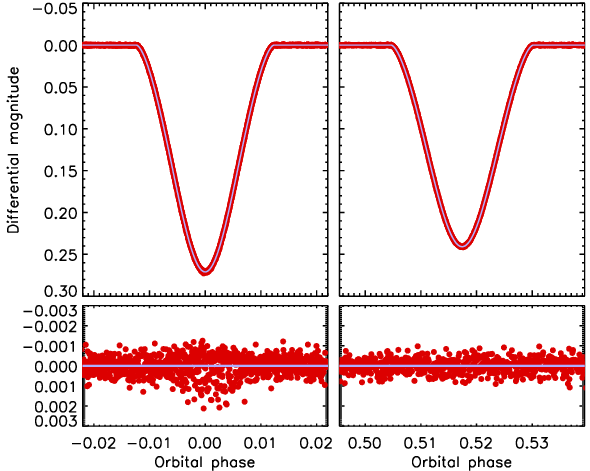} \\
\caption{\label{fig:phase} {\sc jktebop} best fit to the light curves of \targ\ from 
TESS sector 86 for the primary eclipse (left panels) and secondary eclipse (right panels). 
The data are shown as filled red circles and the best fit as a light blue solid line. 
The residuals are shown on an enlarged scale in the lower panels.} \end{figure}

For our analysis we defined star~A to be the star eclipsed at the primary (deeper) minimum, and star~B to be the one eclipsed at secondary minimum. The fitted parameters for each sector were fractional radii of the stars ($r_{\rm A}$ and $r_{\rm B}$), the central surface brightness ratio ($J$), third light ($L_3$), orbital inclination ($i$) and eccentricity ($e$), argument of periastron ($\omega$), orbital period ($P$), and a reference time of primary minimum ($T_0$). The fractional radii were expressed as their sum ($r_{\rm A}+r_{\rm B}$) and ratio ($k = {r_{\rm B}}/{r_{\rm A}}$), and the shape parameters as the combinations $e\cos\omega$ and $e\sin\omega$, in both cases to decrease correlations between parameters. Limb darkening (LD) was accounted for using the power-2 law \cite{Hestroffer97aa,Maxted18aa,Me23obs2} and we required both stars to have the same LD coefficients. The linear coefficient ($c$) was fitted and the non-linear coefficient ($\alpha$) fixed at a theoretical value \cite{ClaretSouthworth22aa,ClaretSouthworth23aa}. The TESS flux measurement errors were scaled to force a reduced $\chi^2$ of $\chir = 1.0$. 

\begin{table} \centering
\caption{\em \label{tab:jktebop} Photometric parameters of \targ\ measured using 
{\sc jktebop} from the TESS light curves. The errorbars are 1$\sigma$ standard 
errors and were obtained from the scatter of the results for individual sectors.}
\begin{tabular}{lcc}
{\em Parameter}                           &              {\em Value}            \\[3pt]
{\it Fitted parameters:} \\
Orbital inclination (\degr)               & $      87.5554     \pm  0.0060    $ \\
Sum of the fractional radii               & $       0.08940    \pm  0.00006   $ \\
Ratio of the radii                        & $       0.9220     \pm  0.0071    $ \\
Central surface brightness ratio          & $       0.95684    \pm  0.00084   $ \\
Third light                               & $       0.0049     \pm  0.0020    $ \\
$e\cos\omega$                             & $       0.027355   \pm  0.000005  $ \\
$e\sin\omega$                             & $       0.02411    \pm  0.00016   $ \\
LD coefficient $c$                        & $       0.6208     \pm  0.0065    $ \\
LD coefficient $\alpha$                   &            0.5548 (fixed)           \\
{\it Derived parameters:} \\
Fractional radius of star~A               & $       0.04652    \pm  0.00017   $ \\       
Fractional radius of star~B               & $       0.04288    \pm  0.00018   $ \\       
Light ratio $\ell_{\rm B}/\ell_{\rm A}$   & $       0.814      \pm  0.013     $ \\[3pt]
Orbital eccentricity                      & $       0.03647    \pm  0.00011   $ \\
Argument of periastron ($^\circ$)         & $      41.38       \pm  0.19      $ \\
\end{tabular}
\end{table}

We found good fits for all sectors, and that for sector 86 can be seen in Fig.~\ref{fig:phase}. The results between sectors are also in good agreement. The unweighted mean and standard error of the values for each parameter can be found in Table~\ref{tab:jktebop}. Uncertainties were also calculated using Monte Carlo and residual-permutation algorithms, with similar results to the standard errors in Table~\ref{tab:jktebop}.

Our results for some parameters ($i$, $e$, $\omega$) are in good agreement with those that can easily be compared to the values from MKO10 (their table~4). We find a much smaller third light of 0.5\% compared to their $\sim$10\%, and this changes the measured radii significantly.


\section*{Orbital ephemeris}

\begin{table} \centering
\caption{\em Times of published mid-eclipse for \targ\ and their residuals versus the fitted ephemeris.\label{tab:tmin}}
\setlength{\tabcolsep}{10pt}
\begin{tabular}{rrrrr}
{\em Orbital} & {\em Eclipse time}  & {\em Uncertainty} & {\em Residual} & {\em TESS}   \\
{\em cycle}   & {\em (BJD$_{TDB}$)} & {\em (d)}         & {\em (d)}      & {\em sector} \\[3pt]
   $-$102     &   2458692.937603    &     0.000036      &  $ $0.000001   &      14      \\             
    $-$99     &   2458725.222208    &     0.000022      &  $-$0.000024   &      15      \\             
    $-$72     &   2459015.783904    &     0.000028      &  $-$0.000004   &      26      \\             
    $-$36     &   2459403.199411    &     0.000024      &  $-$0.000064   &      40      \\             
    $-$33     &   2459435.484155    &     0.000017      &  $ $0.000049   &      41      \\             
     $-$3     &   2459758.330385    &     0.000016      &  $-$0.000027   &      53      \\             
        0     &   2459790.615020    &     0.000021      &  $-$0.000023   &      54      \\             
        2     &   2459812.138137    &     0.000006      &  $ $0.000007   &      55      \\             
       12     &   2459919.753581    &     0.000047      &  $ $0.000016   &      59      \\             
       50     &   2460328.692216    &     0.000024      &  $-$0.000004   &      74      \\             
       53     &   2460360.976867    &     0.000022      &  $ $0.000016   &      75      \\             
       65     &   2460490.115396    &     0.000021      &  $ $0.000023   &      80      \\             
       70     &   2460543.923052    &     0.000024      &  $-$0.000039   &      82      \\             
       80     &   2460651.538493    &     0.000024      &  $-$0.000033   &      86      \\             
\end{tabular}
\end{table}

\begin{figure}[t] \centering \includegraphics[width=\textwidth]{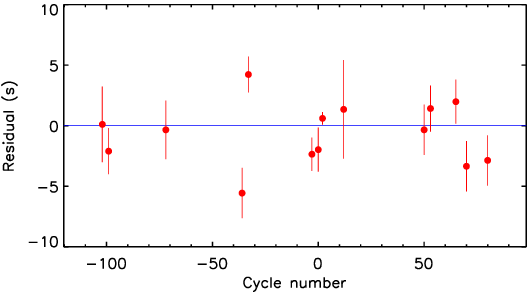} \\
\caption{\label{fig:tmin} Residuals of the times of minimum light from 
Table~\ref{tab:tmin} (red circles) versus the best-fitting ephemeris. 
The blue solid line indicates a residual of zero.} \end{figure}

Our analysis above yielded a mean time of primary eclipse for each sector, which are useful for establishing a precise ephemeris. We fitted a linear ephemeris to the times, obtaining 
\begin{equation} \label{eq:ephem}
  \mbox{Min~I} = {\rm BJD}_{\rm TDB}~ 2459790.615043 (6) + 10.76154354 (16) E
\end{equation}
where $E$ is the number of cycles since the reference time of minimum and the bracketed quantities indicate the uncertainty in the final digit of the previous number. The root-mean-square of the residuals is 5.2~s and the $\chir$ is 2.0. This relatively poor agreement may be caused by spot activity on the stars affecting the eclipse shapes (see below). The uncertainties in the ephemeris in Eq.~\ref{eq:ephem} have been multiplied by $\sqrt{\chir}$ to account for this. The times of minimum are given in Table~\ref{tab:tmin} and the ephemeris is plotted in Fig.~\ref{fig:tmin}. The relatively larger uncertainty in the timing from sector 59 is because there was only one fully-observed primary eclipse in these data.

We fitted a quadratic ephemeris as well, to see if apsidal motion was detectable, but the fit showed a negligible improvement and the quadratic variation was not significantly detected. We also tried to include historical times of minimum but found that they do not match the ephemeris above, suggesting the possibility of a light-time effect from a third body. We leave this matter to future work.


\section*{Radial velocity analysis}

\begin{figure}[t] \centering \includegraphics[width=\textwidth]{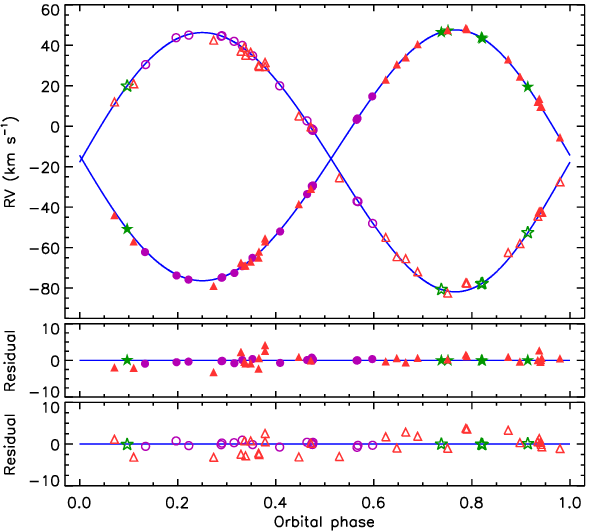} \\
\caption{\label{fig:rv} RVs of \targ\ from MKO10 compared to the best fit from 
{\sc jktebop} (solid blue lines). The RVs for star~A are shown with filled symbols, 
and for star~B with open symbols. The residuals are given in the lower panels 
separately for the two components. RVs from CORAVEL are shown with purple circles, 
from \'Elodie with green stars, and from Asiago with red triangles.} \end{figure}

We have reanalysed the RVs presented by MKO10 for two reasons: to provide an independent analysis; and to obtain the velocity amplitudes ($K_{\rm A}$ and $K_{\rm B}$) needed for the next section (below). MKO10 performed a joint fit of their light curves and the RVs, and proceeded directly to the masses and radii without passing through the intermediate quantities $K_{\rm A}$ and $K_{\rm B}$. 

The RVs in MKO10 comprise three sets: 17 RVs per star from the CORAVEL cross-correlation spectrometer \cite{Baranne++79va}, six spectra from the \'Elodie spectrograph \cite{Baranne+96aas} giving six RVs for star~A and five for star~B, and 29 spectra from the Asiago \'echelle spectrograph giving one RV per star per spectrum. The Asiago RVs were already published in Milone et al.\ \cite{Milone+05aa}. We included all RVs in a single analysis, except for rejecting one discrepant Asiago RV taken near secondary eclipse.

We fitted all RVs simultaneously using {\sc jktebop} with a fixed $P$ but allowing for a shift in $T_0$. We also fitted for $K_{\rm A}$ and $K_{\rm B}$, the systemic velocity for both stars separately, $e\cos\omega$, and $e\sin\omega$. We also tried alternative approaches with $e\cos\omega$ and $e\sin\omega$ fixed at the photometric values and/or forcing the systemic velocities of the two stars to be the same, with essentially the same results but smaller errorbars. The outcome of this analysis is the measurements $K_{\rm A} = 61.971 \pm 0.056$\kms, $K_{\rm B} = 64.067 \pm 0.060$\kms, and the plot in Fig.~\ref{fig:rv}. The RV fit yielded an insignificant phase shift versus our orbital ephemeris, and $e\cos\omega$ and $e\sin\omega$ consistent with the light curve analysis. The errorbars were obtained from 1000 Monte Carlo simulations \cite{Me21obs5}. 

%
%


\section*{Physical properties and distance to \targ}

\begin{table} \centering
\caption{\em Physical properties of \targ\ defined using the nominal solar units 
given by IAU 2015 Resolution B3 (ref.~\cite{Prsa+16aj}). \label{tab:absdim}}
\begin{tabular}{lr@{~$\pm$~}lr@{~$\pm$~}l}
{\em Parameter}        & \multicolumn{2}{c}{\em Star A} & \multicolumn{2}{c}{\em Star B}    \\[3pt]
Mass ratio   $M_{\rm B}/M_{\rm A}$          & \multicolumn{4}{c}{$0.9673 \pm 0.0013$}       \\
Semimajor axis of relative orbit (\Rsunnom) & \multicolumn{4}{c}{$26.814 \pm 0.017$}        \\
Mass (\Msunnom)                             &  1.1354 & 0.0023      &  1.0984 & 0.0022      \\
Radius (\Rsunnom)                           &  1.2474 & 0.0046      &  1.1498 & 0.0049      \\
Surface gravity ($\log$[cgs])               &  4.3012 & 0.0032      &  4.3576 & 0.0037      \\
Density ($\!\!$\rhosun)                     &  0.5850 & 0.0064      &  0.7226 & 0.0091      \\
Synchronous rotational velocity ($\!\!$\kms)&  5.864  & 0.021       &  5.406  & 0.023       \\
Effective temperature (K)                   &  6000   & 150         &  5935   & 150         \\
Luminosity $\log(L/\Lsunnom)$               &  0.259  & 0.044       &  0.170  & 0.033       \\
$M_{\rm bol}$ (mag)                         &  4.09   & 0.11        &  4.32   & 0.11        \\
Interstellar reddening \EBV\ (mag)          & \multicolumn{4}{c}{$0.00 \pm 0.01$}			\\
Distance (pc)                               & \multicolumn{4}{c}{$77.9 \pm 1.2$}            \\[3pt]
\end{tabular}
\end{table}


We calculated the physical properties of \targ\ using the {\sc jktabsdim} code \cite{Me++05aa} with the photometric properties from Table~\ref{tab:jktebop}, and the $K_{\rm A}$ and $K_{\rm B}$ found above. We adopted the \Teff\ of star~A to be $6000 \pm 150$~K from MKO10. For star~B we calculated our own value based on that for star~A and the surface brightness ratio (Table~\ref{tab:jktebop}). The results are given in Table~\ref{tab:absdim}. Based on the data available, we have been able to measure the masses of the stars to 0.2\% precision and the radii to 0.4\% precision. These are among the most precise measurements currently available \cite{Maxted+20mn,Me15aspc}. Compared to MKO10 we find almost identical masses, as expected, but significantly different radii (1.247\Rsun\ and 1.150\Rsun\ versus 1.371\Rsun\ and 1.052\Rsun). We attribute this discrepancy to the much greater quality and quantity of the TESS light curves compared to previous ground-based datasets, and to our somewhat different photometric solution with much less third light.



We determined the distance to \targ\ using the $BV$ magnitudes from Tycho \cite{Hog+00aa}, $JHK_s$ magnitudes from 2MASS \cite{Cutri+03book} and the surface brightness calibrations from Kervella et al.\ \cite{Kervella+04aa}. No interstellar reddening was needed to align the optical and infrared distances, but we allowed an uncertainty of 0.01~mag for this deduction. Our most precise distance estimate is in the $K_s$ band and is $77.9 \pm 1.2$~pc; this is consistent with the distance of $79.27 \pm 0.32$~pc from inversion of the \gaia\ DR3 parallax \cite{Gaia23aa}.

We performed a comparison of the measured masses, radii and \Teff\ value of the component stars to theoretical predictions to infer their age and chemical composition. We found that a {\sc parsec} 1.2 theoretical model \cite{Bressan+12mn} with an approximately solar chemical composition and an age around 3.5~Gyr provides an adequate match to our results. We leave detailed analysis to the future, preferably once a spectroscopic metallicity estimate is available to provide another constraint on the theoretical models.


\section*{Stellar activity}


\begin{figure}[t] \centering \includegraphics[width=\textwidth]{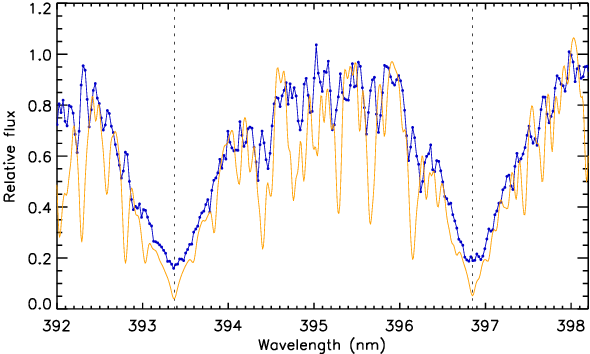} \\
\caption{\label{fig:cahk} Observed spectrum of \targ\ around the Ca~{\sc ii} H and K 
lines (blue line line with points) compared to a synthetic spectrum for a star with 
$\Teff = 6000$~K, $\logg = 4.2$ and solar metallicity from the BT-Settl model atmospheres 
\cite{Allard+01apj,Allard++12rspta} (orange line). The H and K line central wavelengths 
are shown with dotted lines. The spectrum of \targ\ has been shifted to zero velocity 
and normalised to approximately unit flux.} \end{figure}

The TESS light curve shows minor brightness modulations due to starspot activity, in addition to the instrumental variations discussed above. There are hints of a recurrence period of 11.0 to 11.5~d, in which case at least one of the stars is rotating slightly slower than the orbital period. The spot modulation evolves on a similar timescale so this rotation period remains tentative. In a previous analysis of ZZ~UMa \cite{Me22obs6} we found spot activity to be accompanied by a gradual change in the light ratio of the stars; this was searched for but not significantly detected in the current case.

In order to investigate the possibility of magnetic activity, we observed the Ca~{\sc ii} H and K lines of \targ\ using the Intermediate Dispersion Spectrograph (IDS) at the Cassegrain focus of the Isaac Newton Telescope (INT). A single observation with an exposure time of 150~s was obtained on the night of 2022/06/07 in excellent weather conditions. We used the 235~mm camera, H2400B grating, EEV10 CCD and a 1~arcsec slit and obtained a resolution of approximately 0.05~nm. A central wavelength of 4050\,\AA\ yielded a spectrum covering 373--438~nm at a reciprocal dispersion of 0.023~nm~px$^{-1}$. The data were reduced using a pipeline currently being written by the author \cite{Me+20mn2}, which performs bias subtraction, division by a flat-field from a tungsten lamp, aperture extraction, and wavelength calibration using copper-argon and copper-neon arc lamp spectra.

The spectrum was obtained at orbital phase 0.831 and is compared in Fig.~\ref{fig:cahk} to a synthetic spectrum without chromospheric activity \cite{Allard+01apj,Allard++12rspta}. The Ca H and K line centres are clearly filled in by emission. The two stars had an RV separation of 118\kms\ at this time (0.156~nm at 396.85~nm) and double-peaked emission with this separation is apparent. We conclude that both stars show chromospheric emission due to spot activity, and spot modulation of at least one star is visible in the TESS light curves.

\section*{Summary and conclusions}

\targ\ is dEB containing two late-F stars in a eccentric orbit with a relatively long period \cite{Me24obs3} of 10.76~d. It has a white-dwarf companion at an angular separation of 14.4\as\ and a hint of eclipse timing variations caused by another, closer, companion. The TESS mission has observed it in 14 sectors, with full coverage of 63 eclipses (31 primary and 32 secondary). We modelled these light curves and published RVs to measure the physical properties of the system. These properties are matched by theoretical predictions for an age of 3.5~Gyr and an approximately solar metallicity. The distance we find agrees with that from \gaia\ DR3. Both stars exhibit chromospheric activity in the Ca H and K lines.

\targ\ would benefit from a detailed spectroscopic analysis to determine the photospheric chemical composition and improve measurements of the stellar \Teff\ values. An analysis of its times of minimum light would also be helpful in checking for apsidal motion and the existence of a third body (in addition to the white dwarf). As the stellar masses are measured to 0.2\% and the radii to 0.4\%, and even without the suggested future work, the components of \targ\ are now among the most precisely characterised stars known.


\section*{Acknowledgements}

This paper includes data collected by the TESS\ mission and obtained from the MAST data archive at the Space Telescope Science Institute (STScI). Funding for the TESS\ mission is provided by the NASA's Science Mission Directorate. STScI is operated by the Association of Universities for Research in Astronomy, Inc., under NASA contract NAS 5–26555.
This paper includes observations made with the Isaac Newton Telescope operated on the island of La Palma by the Isaac Newton Group of Telescopes in the Spanish Observatorio del Roque de los Muchachos of the Instituto de Astrof\'{\i}sica de Canarias.
This work has made use of data from the European Space Agency (ESA) mission {\it Gaia}\footnote{\texttt{https://www.cosmos.esa.int/gaia}}, processed by the {\it Gaia} Data Processing and Analysis Consortium (DPAC\footnote{\texttt{https://www.cosmos.esa.int/web/gaia/dpac/consortium}}). Funding for the DPAC has been provided by national institutions, in particular the institutions participating in the {\it Gaia} Multilateral Agreement.
The following resources were used in the course of this work: the NASA Astrophysics Data System; the SIMBAD database operated at CDS, Strasbourg, France; and the ar$\chi$iv scientific paper preprint service operated by Cornell University.



\end{document}